\documentclass[10pt]{iopart} 
\usepackage{psfig,epsfig,color} 
\begin{document}

\title{Modelling tubular shapes in the inner mitochondrial membrane}
\author{A. Ponnuswamy$^1$, J. Nulton$^2$, J. M. Mahaffy$^2$,\\ P. Salamon$^2$,
T. G. Frey$^3$,  and A. R. C. Baljon$^1$}
\address{$^1$Department of Physics, San Diego State University, CA 92182,\\
$^2$Department of Mathematics, San Diego State University, CA 92182,\\
$^3$Department of Biology, San Diego State University, CA 92182 }

\begin{abstract}  
The inner mitochondrial membrane has been shown to have a novel
structure that contains tubular components whose radii are on the order of
10 nm as well as comparatively flat regions~\cite{frey2000}. 
The structural organization of mitochondria is important
to understanding their functionality. We present a model that can account, thermodynamically,
for the observed size of the tubules.  The model  contains two
lipid constituents with different shapes.  They are allowed to distribute 
in such a way that the composition differs on the two sides of the tubular
membrane.                       
Our calculations make two predictions: (1) there is a pressure
difference of 0.2 atmospheres across the inner membrane as a necessary
consequence of the experimentally observed tubule radius of 10 nm. and (2)
migration of differently shaped lipids causes concentration variations 
between the two sides of the tubular membrane on
the order of 7 percent.  

\end{abstract}
\maketitle
\nopagebreak
\section{Introduction}
Mitochondria are organelles in eukaryotic cells that provide most of the
chemical energy, ATP, from  oxidative metabolism and more
recently have been shown to play a key role in apoptosis or
programmed cell death. Mitochondria have their own DNA and are
thought to have evolved from a prokaryotic organism
that became engulfed and lived inside the ancient eukaryotic
cell. Mitochondria have an outer membrane that surrounds a
complex inner membrane structure that in turn encloses the matrix
space of the organelle. The inner membrane and the matrix
contain a rich collection of enzymes that are crucial to
breaking down a number of metabolites such as fatty acids and pyruvate forming Acetyl-CoA.  Acetyl-CoA is oxidized via the Citric Acid Cycle to produce reduced nucleotides, NADH and FADH2, that provide reducing potential for the mitochondrial electron transport chain that converts this energy into an electrochemical proton gradient across the inner mitochondrial membrane that the ATP Synthase uses to synthesize ATP, the principal source of energy for cell function.
Mitochondria  also interact with the cell in the
process of apoptosis. When mitochondria receive certain 
signals  they undergo a structural transformation that leads to the
release of cytochrome c, which in turn causes the cell to 
destroy itself in a controlled process.

Our research examines the physical structure of mitochondria in
hopes of better  understanding a number of the key functions
performed by this organelle. 
Electron tomography has provided high resolution three-dimensional structures 
of ``orthodox" (healthy) mitochondria {\em in vivo}~\cite{frey2000,alberts,perkins2000}.
\begin{figure}
\setlength{\unitlength}{1mm}
\begin{picture}(100,100)(0,0)
\put(0,0){\epsfxsize=15cm\epsfbox{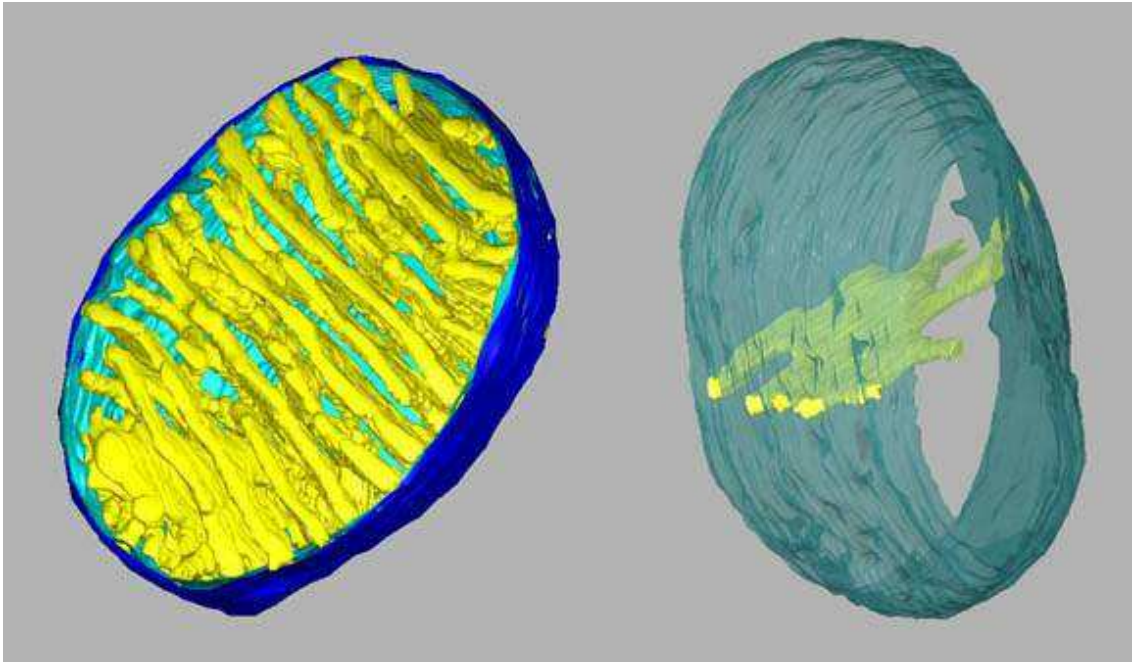}}
\end{picture}
\caption[]{Figure 1:  3D computer models of the the mitochondrial membranes
generated from the electron tomogram of a mitochondrion observed in
chick cerebellum prepared by conventional chemical fixation,
dehydration, and embedding techniques.  The image on the left shows
the outer membrane in dark blue, inner boundary membrane in
turquoise, and all of the cristae in yellow.  The image on the right
shows a typical crista in yellow connected to the inner boundary
membrane that has been rendered translucent.  The connections are via
crista junctions of uniform size and tubular cristae segments of
variable length.  Quicktime movies of 3D models of mitochondria
rotating about an axis can be viewed in full color at:
http://www.sci.sdsu.edu/TFrey/MitoMovie.htm
\label {tomo}
}
\end{figure}    
These structures, shown in Figure 1, exhibit several features necessary for proper function.
The inner mitochondrial membrane consists of a
lipid bilayer that has two
components. (1) an inner boundary membrane  (IBM) that lies closely apposed to
the outer membrane and (2) a crista membrane that projects into the matrix
forming cristae, which are either tubular in shape or lamellar.  
Tubules are connected to the IBM by crista junctions (see Fig. 1), shaped like the bell of a trumpet~\cite{renken2002}.
The observed mitochondria have a large matrix volume that pushes the
inner boundary membrane against the outer membrane and collapses the cristae
into flat lamellar compartments.

In our work, we explore the possibility that at least portions of the mitochondrial membrane make up a thermodynamically stable structure that minimizes free energy~\cite{renken2002}.
Rather than try to deduce the morphology from first principles~\cite{handbook1995},
 we take the observed morphology as given and make inferences regarding the 
physico-chemical environment in which this morphology could exist.

We begin by noting that the observed morphology shows a definite scale.  
That the crista junctions have a fairly constant radius of about 10 nm has 
been noted in several places~\cite{frey2000,perkins2000,renken2002}.  
In fact, their diameter roughly matches the spacing between the lamellar 
regions of the cristae and that of the tubules linking these regions 
to the junctions.  It is certainly possible that some skeletal components maintain the spacing everywhere and thereby account for the scale. 
We consider the hypothesis that such skeletal elements exist only in the 
lamellae whose surfaces house the machinery of ATP production which probably 
requires (and gives) some mechanical stability at a spacing that roughly 
matches the distance that the membrane bound proteins extend into the 
intermembrane space.  In that case the shape of the tubular regions is 
determined by elastic energy minimization rather than skeletal elements. 

Suppose that a cylindrical tubular bilayer of fixed length is constrained in 
such a way that it can only increase or decrease its radius by exchanging 
area (molecules) with a flat membrane as a reservoir.  
It is not surprising, all other things being equal, that its radius would 
grow indefinitely in order to mitigate the energetic cost of bending 
required by the formation of the cylindrical tubule.  
If, however, there were a positive osmotic pressure difference across 
the membrane favoring the exterior of the tube, 
i.e. the mitochondrial matrix, then osmotic work would be required to grow 
the radius of the tube.  The result is a tradeoff of two energetic components 
(bending and pressure work), giving an equilibrium tube radius whose magnitude 
depends on the pressure difference.  Although such a pressure difference has 
not been measured, the matrix volume has been shown to respond to changes 
in osmolarity of the surrounding media~\cite{hackenbrock1968}, and the crista junction diameters 
respond to changes in matrix volume~\cite{renken2004}

The IMM has been shown to contain several types of phospholipids. In addition, 50\% of membrane surface is occupied by proteins, while proteins make up approximately 75\% of the inner membrane mass.  For the sake of simplicity, our model includes no proteins and only the two most common lipid types: phosphatidyl ethanolamine (PE) and phosphatidyl choline (PC). These occur naturally in the IMM at fractions of 27.7\% and 44.5\%, respectively.  Moreover, in our model, we consider only the dioleic acid esters of the lipds, DOPE and DOPC, each of which is heterogeneous with respect to its fatty acid composition.  Although most authors neglect 
membrane composition 
altogether, some have attributed the variations in membrane curvature to the existence of domains of  
differently shaped molecules~
\cite{godzd1998,julich1993}.
While for lipids of limited viscibility~\cite{baumgart2003} these can be seen, we do 
not expect this to be the case for DOPE and DOPC which are chemically very 
similar and thus should form nearly ideal solutions in which the entropic 
incentive to mix is far outweighed by possible energetic advantages of 
segregation. 
  However,  a different lipid composition on in the two 
monolayers of the tubular membrane is to be expected.  Hence, we  
assess the extent to which the geometry of the lipids contributes 
to the shape of the membrane.
The contest here is between the entropic contribution to the free energy and the bending energy savings obtained by distributing the molecules according to shape.

We formulate the free energy of a tubule plus surroundings as a function of  its
radius and composition.  Optimality with respect to variation 
of the radius gives a predicted osmotic pressure difference $\Delta p$ across 
the membrane.  Optimality with respect to composition predicts the extent to 
which shape-based redistribution takes place among the molecules.  The two most 
extreme curvature environments are represented by the inner and outer monolayers 
of the tube.  The composition of the principal lipid is calculated to vary by 
about 7\% between these two regions for the observed tubular size.  This result reveals a dominant role 
played by the entropic contribution to the free energy at normal physiological 
temperatures.

Although our approach does not come close to explaining all aspects of inner 
membrane morphology, it is well grounded in experimental observations and 
enables us to leverage observed morphologies into predictions regarding 
additional aspects of the physico-chemical environment in which membrane 
morphology is observed.

\section{Formulation of the free energy}

In this section we formulate the free energy of a tubule  and its surroundings as
a function of its radius and composition. 
The flat portions act as a reservoir which constrains 
the chemical potential of the lipid 
molecules in the tubules, which, by our assumption, must be in 
equilibrium with this reservoir. Since the reservoir is a bilayer
comprising a surface of mean curvature zero, the lipid compositions on the 
two sides (inner and outer) are the same, at least as far as bending forces 
are concerned. Short of postulating a preference of some lipids for the 
chemical environment on the two sides of the membrane, we may assume that 
the compositions on the two sides are the same and act as a reservoir 
for lipid molecules in the tubular regions.
Hence, we consider $N_E^{(*)}$ molecules of DOPE and $N_C^{(*)}$ molecules
of DOPC distributed among the inner and outer layers of a
cylindrical bilayer of unit length and a flat bilayer
reservoir.  Let $U$ and $S$ denote respectively the total bending
energy and the total entropy of the membrane molecules, and
$T$ the temperature.  
Then, up to constants in the radius and the compositions of the
membrane, the sum of the free energies of all the systems that
participate in the energetics of altering the radius or the
tubule and its composition can be written as

\begin{equation}
\label{1}
G=U-TS+ \Delta pV.
\end{equation}

\noindent In this equation we have dropped the $pV$ term for the
membrane and the $U$ and $TS$ contributions of the
surrounding cytosol. Thus the $U-T S$ portion is the free
energy of the membrane while the
$\Delta pV$ term is the free energy of the matrix and
intermembrane region, which depends on the volume V inside the cylindrical tubule and the osmotic pressure difference $\Delta p$  between the matrix and intermembrane space, with the higher pressure in the matrix.
We will use the following notational
conventions. Subscripts $E$ and $C$ will continue to denote
the molecular species DOPE and DOPC.  Superscripts $(i)$,
$(o)$, and $(r)$ will refer, respectively, to the inner
monolayer, the outer monolayer of the tubule, and 
either monolayer of the flat bilayer reservoir.  $N$ will
continue to indicate the number of
molecules, and lower case letters
$u$ and $s$ will indicate energy and
entropy per molecule.  More precisely,
$s$ will denote a partial molecular
entropy.  For example, $u_E^{(o)}$ is the partial molecular
bending energy of DOPE
residing in the outer monolayer of the
cylindrical tubule.

Consider first the total bending energy $U$.  The 
conventional approach is to employ Helfrich's theory~\cite{helfrich1973} and to take the free energy density per unit membrane area as

\begin{equation}
\label{25}
{\it U}=\frac{1}{2}{\kappa_b}
\left(C-{C_s}\right)^2,
\end{equation}

\noindent where $C$ and $C_s$ are the ambient and sponteneous curvatures of the membrane and $\kappa_b$ is the bending modulus.
However, to allow us to study the lipid redistribution between the monolayers 
of the tubular membrane and the reservoir,
we employ a molecular level model.
Our model takes the bending energy of the bilayer to be additive over the 
individual lipids in each of the monolayers.  
Following Israelachvilli  ~\cite{israe1995}, we take the bending energy of one lipid molecule in the cylindrical monolayer to be
\begin{equation}
\label{24}
u=\frac{1}{2}{K_A}
a\left(1-\frac{a_s}{a}\right)^2,
\end{equation}
where $K_A$ is the 
compressibility modulus,  and $a$ is the characteristic
interfacial area of the lipid
at the ambient curvature $C$.  The compressibility modulus is related to the 
bending modulus.  For small deformations $\kappa_b$ depends linearly on 
$K_A$ and the square of the membrane thickness
~\cite{landau1986,boal2002}.
The relation between $a$ and $C$ will be given
shortly.                         
For a monolayer containing one type of lipid, $a$ 
is defined as the area of the membrane divided by the total number of lipids. 
$a_s$ is the  characteristic interfacial area for a monolayer at the spontaneous 
curvature. Since the spontaneous curvature and hence $a_s$ depend 
on the type of lipid, the bending energy differs as well.  
This causes a redistribution over the two leaflets, since they both have 
different ambient curvatures. 
Spontaneous curvatures have been determined
experimentally and are understood as the curvature
of ``choice" for a particular lipid type constrained to a cylindrical monolayer
with minimum
bending energy.

The total bending energy 
\begin{eqnarray}
\label{2}
U=N_E^{(i)}u_E^{(i)}+N_C^{(i)}u_C^{(i)}+
N_E^{(o)}u_E^{(o)}+N_C^{(o)}u_C^{(o)}
\nonumber\\
+(N_E^{(*)}-N_E^{(i)}-N_E^{(o)})u_E^{(r)}+
(N_C^{(*)}-N_C^{(i)}-N_C^{(o)})u_C^{(r)}.
\end{eqnarray}
These terms can be rearranged to give
\begin{equation}
\label{3}
U=N_E^{(i)}\Delta u_E^{(i)}+
N_C^{(i)}\Delta u_C^{(i)}+
N_E^{(o)}\Delta u_E^{(o)}+
N_C^{(o)}\Delta u_C^{(o)}+U^{(*)},
\end{equation}
where,  all  four $\Delta's$ are defined relative to the value of each 
quantity in the reservoir.  For example
\begin{equation}
\label{4}
\Delta u_C^{(o)}=u_C^{(o)}-u_C^{(r)}.
\end{equation}
We have also defined the quantity
\begin{equation}
\label{5}
U^{(*)}=N_E^{(*)}u_E^{(r)}+N_C^{(*)}u_C^{(r)},
\end{equation}
which does not vary as the molecules are
redistributed among the three compartments.
We introduce $\alpha$ and $\beta$ to
represent the fraction of
DOPE on the inner and the outer monolayers
of the tubule, respectively.  With this, $U$ can be
rewritten:
\begin{equation}
\label{6}
U=N^{(i)}(\alpha\Delta
u_E^{(i)}+(1-\alpha)
\Delta u_C^{(i)})+
N^{(o)}(\beta\Delta
u_E^{(o)}+(1-\beta)
\Delta u_C^{(o)})+U^{(*)}.
\end{equation}

\noindent A perfectly analogous formula holds for $S$:
\begin{equation}
\label{7}
S=N^{(i)}(\alpha\Delta
s_E^{(i)}+(1-\alpha)
\Delta s_C^{(i)})+
N^{(o)}(\beta\Delta
s_E^{(o)}+(1-\beta)
\Delta s_C^{(o)})+S^{(*)}.
\end{equation}

It remains to formulate the $\Delta
s$'s, the $N$'s, and the
$\Delta u$'s.
The $\Delta s$'s depend not only on the
fractions $\alpha$ and $\beta$, but also on
the fraction $\gamma$ of DOPE in the flat reservoir.
Assuming that the membrane contains only DOPE and DOPC and that their ratio is 
that of the ratio of PE and PC in mitochondrial inner membranes, we have taken
$\gamma=\frac{27.7}{27.7+44.5}=0.384$. We define
$\Delta s_E^{(i)}$ and analogous terms to be $s_E^{(i)}-s_E^{(r)}$.  The partial molecular entropies are
decomposed into a pure part and a mixing part.  Since the
pure part is the same in all three compartments $(i,o,r)$,
the
$\Delta s$'s depend only on the mixing, {\em i.e.}, we write
$\Delta s_E^{(i)}=s_{E, mix}^{(i)}-s_{E,
mix}^{(r)}$.

The total entropy of (ideal)
mixing of the two species on the inner
monolayer of the tubule is given by:
\begin{equation}
\label{8}
S_{mix}^{(i)}=-k_BN^{(i)}(\alpha\ln\alpha+
(1-\alpha)\ln(1-\alpha)),
\end{equation}
or, more explicitly, as a function of the
molecule numbers:
\begin{equation}
\label{9}
S_{mix}^{(i)})
=-k_B\left[N_E^{(i)}\ln(\frac{N_E^{(i)}}
{N_E^{(i)}+N_C^{(i)}})+
N_C^{(i)}\ln(\frac{N_C^{(i)}}
{N_E^{(i)}+N_C^{(i)}})\right].
\end{equation}
The partial molecular
entropy is obtained as
\begin{equation}
\label{10}
s_{E,
mix}^{(i)}=\frac{\partial
S_{mix}^{(i)}}{\partial N_E^{(i)}}
=-k_B\ln\frac{N_E^{(i)}}
{N_E^{(i)}+N_C^{(i)}}=-k_B\ln\alpha.
\end{equation}
Similarly, $s_{E,mix}^{(r)}=-k_B\ln\gamma$,
and, finally,
\begin{equation}
\label{11}
\Delta
s_E^{(i)}=-k_B\ln\frac{\alpha}{\gamma}.
\end{equation}
The other $\Delta s$'s are obtained
similarly:
\begin{equation}
\label{12}
\Delta
s_E^{(o)}=-k_B\ln\frac{\beta}{\gamma}.
\end{equation}
\begin{equation}
\label{13}
\Delta
s_C^{(i)}=-k_B\ln\frac{1-\alpha}{1-\gamma}.
\end{equation}
\begin{equation}
\label{14}
\Delta
s_C^{(o)}=-k_B\ln\frac{1-\beta}{1-\gamma}.
\end{equation}
These are now substituted into (\ref{7}).

\begin{figure}                                         
\setlength{\unitlength}{1mm}                           
\begin{picture}(100,100)(0,0)                          
\put(0,0){\epsfxsize=12cm\epsfbox{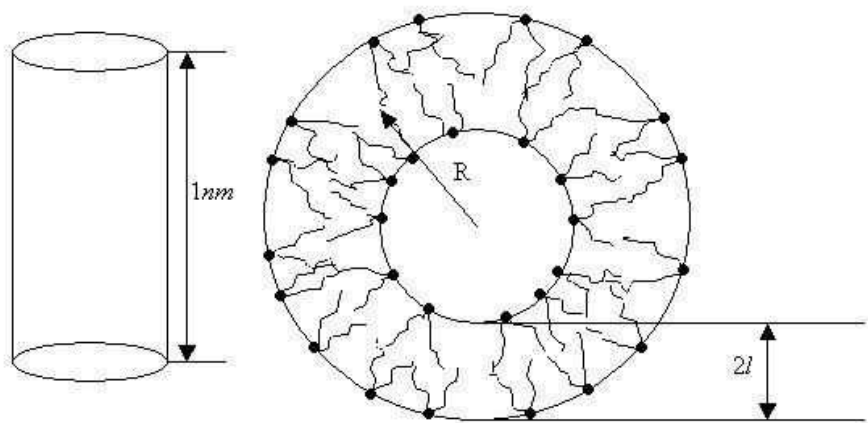}}        
\end{picture}                                                                                               
\caption[]                                                                  
{Sketch of a side view and cross-section of a tubular part of the           
membrane.                                                                   
The radius $R$ is measured from the center of the tubule to the middle of   
the membrane. The thickness of the lipid layer, including only the          
hydrocarbon tails,                                                          
is $2l$, to which each monolayer contributes equally.                       
\label {sketch}                                                             
}                                                                           
\end{figure}                                                                

The interfacial surface of each compartment
($i,o,r$) has its own (cylindrical)
curvature.  They can be expressed in terms of the radius, $R$, of
the midsurface of the cylindrical tubule and the
width, $l$, of the hydrocarbon tails due to one monolayer
(see Figure 2).  It follows that:
\begin{equation}
\label{15}
C^{(i)}=\frac{1}{R-l},
\end{equation}
\begin{equation}
\label{16}
C^{(o)}=\frac{-1}{R+l},
\end{equation}
\begin{equation}
\label{17}
C^{(r)}=0.
\end{equation}

For the immediate discussion we suppress
the superscripts.  Let $v$ be the volume of the hydrocarbon tails of a lipid molecule.  Each molecule
residing in a cylindrical
monolayer with interfacial curvature, $C$,
has a characteristic interfacial area, $a$, which is defined as 
the area of the cylindrical tubule divided by the total number of molecules.
These quantities are related by the
``packing factor" equation:
\begin{equation}
\label{18}
\frac{v}{al}=1+\frac{l}{2}C.
\end{equation}
In our model the hydrocarbon tails of both lipids are identical and hence $v_E=v_C$, so one $a$
fits all molecules in a monolayer.  This
means that the total number of molecules in
a monolayer can be written
\begin{equation}
\label{19}
N=\frac{2\pi}{a|C|}.
\end{equation}
Recall that we take the (fixed) length of
the cylindrical tubule to be 1, for convenience.
Also recall $C^{(o)}$ is negative; hence,
the absolute value.  Combining (\ref{18})
and (\ref{19}), we have

\begin{equation}
\label{20}
N=\frac{2\pi l}{v|C|}(1+\frac{l}{2}C).
\end{equation}
Using (\ref{15}) and (\ref{16}) to adapt
(\ref{20}) to the two cylindrical
monolayers, we have
\begin{equation}
\label{21}
N^{(i)}=\frac{\pi l}{v}(2R-l).
\end{equation}
\begin{equation}
\label{22}
N^{(o)}=\frac{\pi l}{v}(2R+l).
\end{equation}

We now calculate the following bending energies:
\begin{equation}
\label{23}
u_E^{(i)},u_E^{(o)},u_E^{(r)},u_C^{(i)},
u_C^{(o)},u_C^{(r)}.
\end{equation}
Combining (\ref{24}) and
(\ref{18}), we obtain
\begin{equation}
\label{28}
u=\frac{1}{2}K_A
a_s\frac{(C-C_s)^2}{(2/l+C_s)(2/l+C)},
\end{equation}

\noindent Distributing the appropriate subscripts
and superscripts to the quantities
$u$, $K_A$, $C$, and $C_s$, we obtain the
6 quantities (\ref{23}). This completes the formulation of the free energy $G$.

Values for all constants have been obtained from the literature.
The thickness of the layer of hydrocarbon tails, $l$, is assumed to be constant in the model at hand and equals 1.6 nm.~\cite{israe1995}
Experimentally it has been found that, when DOPE and DOPC form
monolayers with a cylindrical shape, their spontaneous
curvatures, $C_{s}$, are the inverse of their intrisic radii of curvature 
respectively $1/2.06$ nm$^{-1}$ and $1/9.05$
nm$^{-1}$~\cite{steicherscott1994}.   The area per headgroup $a_s$ for DOPE
equals 0.163 nm$^2$~\cite{israe1995}.  Using (\ref{18}) for DOPE one
obtains $v=0.362$ nm$^3$.  Since the volume of the hydrocarbon tails of the
two lipid species is the same, one can use the same
formula (\ref{18}) to obtain that $a_s$ for DOPC equals 0.208
nm$^2$.  The compressibility moduli for DOPE and DOPC are $33.0  \ {\rm k}_{B}T/{\rm nm}^{2}$
and $26.4 \ {\rm k}_{B}T/{\rm nm}^{2}$, respectively~\cite{fuller2001}.

\section {Results and Discussion}

\begin{figure}                                                              
\setlength{\unitlength}{1mm}                                                
\begin{picture}(100,100)(0,0)                                               
\put(0,0){\epsfxsize=12cm\epsfbox{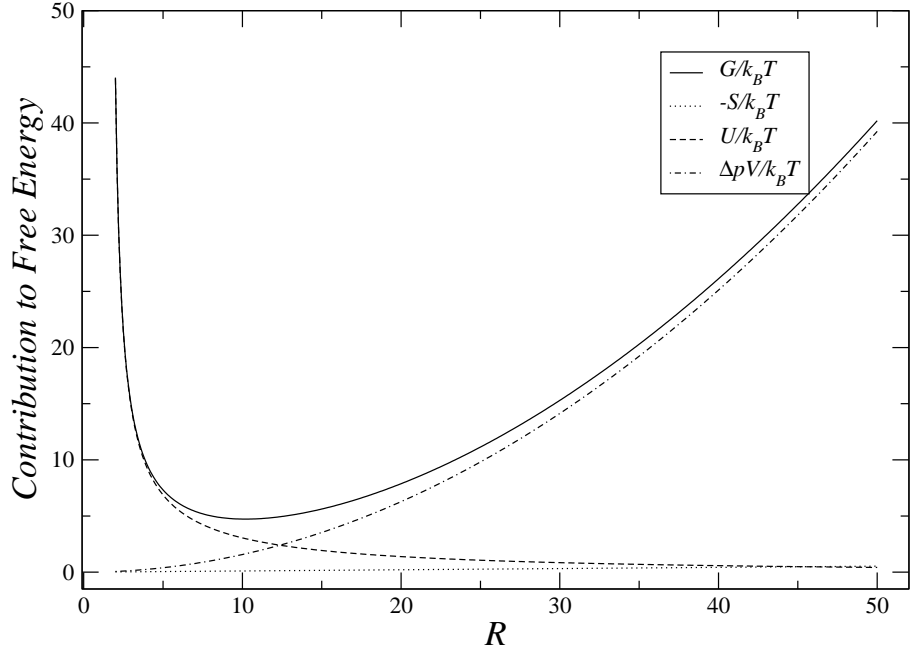}}                            
\end{picture}                                                               
\caption[]                                                                  
{Free energy as function of radius $R$ at $\Delta p =^M                       
 0.005 \ {\rm k}_{B}T/{\rm nm}^{3}$,                                        
$\alpha= 0.423$, and $\beta = 0.352$. The contributions of the different terms
in (\ref{1}) are shown independently.                                       
\label {varyrad}                                                            
}                                                                           
\end{figure}                                                                

For values of $\Delta p$ between 0.4 mbar and 4 bar the free energy 
$G$ as a function of the radius $R$, and the compositions $\alpha$, and $\beta$ has been calculated using 
(\ref{1}).  Figure 3 gives the free energy as a function of $R$
for a pressure difference 
$\Delta p= 0.005 \ {\rm k}_{B}T/{\rm nm}^{3}$ (0.2 bar) and compositions of monolayers of the tubular membrane
given by $\alpha=0.423$ and $\beta=0.352$.   
These values for $\alpha$ and $\beta$ yield the lowest free energy;
changing the compositions of the monolayers
results in a similar graph to that shown in
Figure 3, except that the value of $G/{\rm k}_{B}T$ at which a minimum occurs
is higher.
Figure 3 shows the total free energy as well as the  individual contributions of entropy and bending energy 
of the membrane, and the free energy of the surroundings.  The scale is arbitrary and set so that the free energy vanishes at infinite R, zero  $\Delta p$, and $\alpha=\beta=\gamma$. 
The  pressure difference ($0.2$ bar) is adjusted such that the free energy curve 
has a minimum at $R=9.9$ nm, which is close to the experimentally
observed value.  
The pressure difference correponds to a concentration difference of $8$ mM. 

Setting our scale in Figure 3 so as to make the free energy vanish at infinite R, zero  $\Delta p$, 
and
$\alpha=\beta=\gamma$  amounts to setting 
the quantities $U^{(*)}$ in (\ref{6}) and $S^{(*)}$
in
 (\ref{7}) to zero.  Their values are independent of $R$, $\alpha$, and $\beta$.
Setting them 
to zero will
not change the location of the minimum of free energy.
It follows that all three terms in (\ref{1}) scale linearly
with the length of the tubule.   
Therefore our results are valid for arbitrary length.

\begin{figure}                                                             
\setlength{\unitlength}{1mm}                                               
\begin{picture}(100,100)(0,0)                                              
\put(0,0){\epsfxsize=10cm\epsfbox{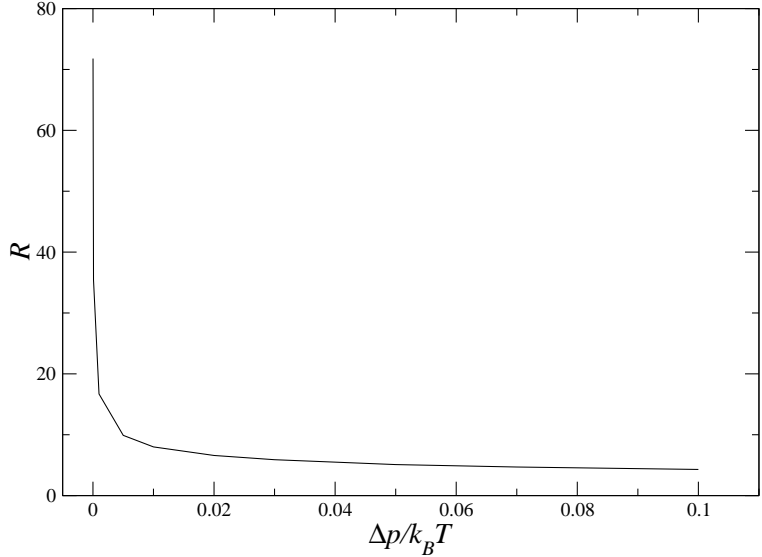}}                          
\end{picture}                                                              
\caption[]                                                                 
{Radius as a function of pressure difference.                              
$\Delta p= 0.025 \ {\rm k}_{B}T/{\rm nm}^{3}$ corresponds to a pressure difference of 1 bar.
\label {pressrad}                                                          
}                                                                          
\end{figure}                                                               

\begin{table} [htb]                                                        
\begin{tabular}{|l|r|r|r|r|r|}                                             
\hline                                                                     
$\Delta p$ (k$_B$T nm$^{-3}$) & $\Delta p$ (bar) & $R$ (nm) &              
G (k$_B$T) & $\alpha$ &                                                    
$\beta$ \\[0.5ex]\hline 0.00001 & 0.0004 & 72.80 & 0.550 &  0.388 &        
0.379\\                                                                    
0.0001 & 0.004 & 35.60 & 1.18 & 0.393 & 0.374\\                            
0.001 & 0.04 & 16.70 & 2.556 & 0.405 & 0.364\\                             
\bf 0.005 & \bf 0.2 & \bf 9.90 & \bf 4.42 & \bf 0.423 & \bf 0.352\\        
0.01 & 0.4 & 8.00 & 5.631 & 0.433 & 0.347\\                                
0.02 & 0.8 & 6.50 & 7.207 & 0.448 & 0.341\\                                
0.03 & 1.2 & 5.80 & 8.358 & 0.458 & 0.336\\                                
0.04 & 1.6 & 5.30 & 9.304 & 0.474 & 0.331\\                                
0.05 & 2.0 & 5.00 & 10.121 & 0.477 & 0.329\\                               
0.07 & 2.8 & 4.60 & 11.526 & 0.486 & 0.327\\                               
0.1 & 4.0 & 4.20 & 13.281 & 0.499 & 0.323\\ \hline                         
\end{tabular}                                                              
\caption[]{Values of the minimum free energy as a function of the pressure difference,
along with the optimum values for $R$, $\alpha$, and $\beta$.}             
\end{table}                                                                
At each value of $\Delta p$, the free energy was minimized and the results
are tabulated in Table 1 which lists values of $R$, $\alpha$, and
$\beta$, that minimize the free
energy $G$, for various values of the pressure difference.  Figure 4 shows
how $R$ varies as a function of $\Delta p$ along this locus of minimum free
energy.  
Interestingly, the radius of 10 nm is reached in the ``elbow" of the curve. 
Increasing the pressure by one order of magnitude decreases the radius by half.
However, decreasing the pressure by one order, increases the radius by a factor of five or so.
As expected, the value of the free energy decreases with increasing radius.  
It will be zero at infinite radius and zero pressure difference.  
At these values, $\alpha=\beta=\gamma$. 
At a finite pressure, more DOPE than average is found in the inner layer and more DOPC in the outer layer. 
At the highest pressure difference $\Delta p= 0.1 \ {\rm k}_{B}T/{\rm nm}^{3}$(4 bar)
the  absolute value of the compositions of the layers in the tubular membrane differ by 17
\%.
 
\begin{figure}                                                             
\setlength{\unitlength}{1mm}                                               
\begin{picture}(100,100)(0,0)                                              
\put(0,0){\epsfxsize=12cm\epsfbox{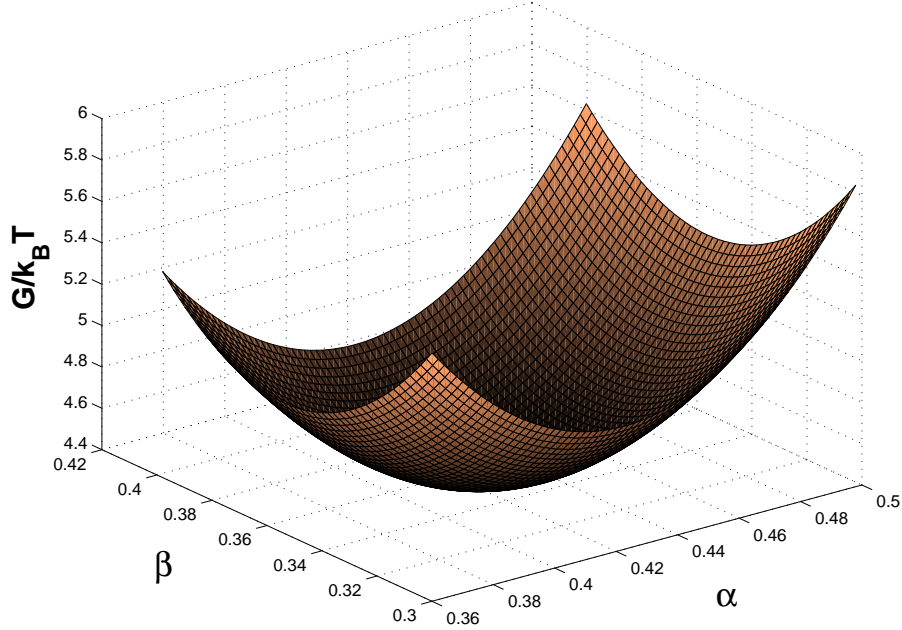}}                             
\end{picture}                                                              
\caption[]                                                                 
{Free energy as a function of $\alpha$ and $\beta$.  The radius of the     
tubule $R=9.9$ nm,  and $\Delta p= 0.005 \ {\rm k}_{B}T/{\rm nm}^{3}$.     
\label {rfix}                                                              
}                                                                          
\end{figure}                                                               

Figure 5 shows the variation of the free energy as a function of lipid 
distribution.  A sharp increase in free energy can be observed when the 
composition deviates from its optimum at $\alpha= 0.423$ 
and $\beta = 0.352$. 

A weakness of the current approach is that the Helfrich energy (\ref{25})
is only valid 
for small deviations from the spontaneous curvature.
Curvatures of the inner and outer monolayers of the tubules differ by up to
100 percent from the spontaneous curvatures. It follows that (\ref{24}) is
only approximately valid.
Currently, we are performing Monte Carlo simulations in hopes of obtaining
the higher-order corrections to this equation.
Using these in the calculations will improve the results as will adding the
effects of other membrane components on the spontaneous curvature and on the
elastic moduli.

\section{Conclusions}
In the present paper we have considered a two-lipid model of the inner
mitochondrial membrane and examined the changes in free energy for the tubular 
parts caused by
variations in shape and
composition. 
The analysis led to two
predictions: (1) The observed radius of 10 nm implies that there is a 0.2 atmosphere osmotic pressure difference
across the
inner membrane  with the higher pressure in the matrix and 
(2) lipids redistribute themselves to give different compositions on the 
two sides of the tubular 
membrane, since the resulting decrease in bending energy is smaller than 
the entropic penalty. 
Using a two lipid model, we found that for crista tubules of the observed size 
the absolute lipid compositions on the two sides of the membrane differ 
by about 7 percent. 
Although the possibility that composition drives shape changes has been 
discussed before~\cite{godzd1998,julich1993}, most approaches in the 
literature neglect a composition dependence.
Without such a dependence, the second term in equation (\ref{1}) is absent and 
the minimum
of the free energy results from a competition between the bending
term and the term due to pressure difference.  Instead of using expression 
(\ref{28}), a Helfrich term (\ref{25}) is then used to model the bending energy. 
As can be seen in Figure 3, in our model the entropic term is almost constant, 
since the composition of the
membrane is more or less uniform.
Hence, as first approximation, it should be possible to express the bending 
energy as a Helfrich term.
The bending energy can then be obtained from a measurement of the bending 
modulus of the inner membrane made on swollen mitoplasts.
The measured value for $\kappa_b$ will yield a value for the bending
energy that 
accounts for all membrane components including
cardiolipin and high concentrations of a variety of integral membrane proteins.

Our model predicts that changes in
the radii of tubules and junctions correspond to variations in
pressure difference.  This might be tested experimentally by manipulating 
the osmotic pressure in preparations of purified mitochondria and observing 
changes in the radii of tubular components.
It has been suggested that the junctions act as a
barrier to the diffusion of cytochrome c.  Indeed Scorrano 
{\em et al.}~\cite{scorrano2002} have observed that
during apoptosis (programmed cell death) the inner boundary membrane remodels, and the 
radii of the tubules increase.  
In certain types of mitochondria, they can increase in these {\em in vitro} 
experiments to 20 nm.
As seen in Table 1,
this corresponds, according to our model, to a large change in the osmotic 
pressure difference.  On the other hand, purified mitochondria that have been 
induced to undergo a permeability transition in buffer of low osmolarity experience an increased $\Delta p$ that causes the matrix to swell.  The crista junctions in these
mitochondria are slightly smaller with radii of 8.5 nm~\cite{renken2004}.

Although the model at hand succesfully describes some of the features of 
the observed morphology, it fails to explain some crucial issues.  
For instance, as can be seen in Figure 3, the minimum value of the free energy for tubules is positive and hence these structures are unstable.   
The tubules will tend to shrink and vanish in the flat membrane regions. 
Additional mechanisms must be at work that prevent them from doing so.  
One  possibility is that such a mechanism is
 provided by proteins and skeletal elements.  
However, we can envision an alternative mechanism.  
Since the inner membrane is confined by an outer one, its area can only grow by 
buckling or by creating protrusions.  
It is very likely that the confinement causes tensile stresses.  
Currently we are investigating the possibility that, thermodynamically,
the combined effects
of osmotic pressure differences and tensile stresses, account for the 
observed  coexistence of  cylindrical tubes of finite radius and 
flat lamellar structures.  

In addition, since the membrane is fluid, tubules will continuously arise, 
grow, shrink, and eventually vanish back into the flat portions 
of the membrane.
It is quite
likely that not just the structural organization of mitochondria,
but also  temporal variations of this structure, are
of importance to understanding mitochondrial functionality.    

\ack  
This research is supported by a grant to AP and ARB from the Donors of the 
Petroleum Research Fund, administered by the American Chemical Society 
and a Blasker Science and Technology Grant from the San Diego Foundation to TGF.  We thank Bjarne Andersen and Christian Renken for helpful conversations.

\end{document}